\documentclass[12pt]{article}
\usepackage{amsmath}
\usepackage{amssymb}

\usepackage{graphicx}

\usepackage[numbers,compress]{natbib}
\begin{document}
\title{Kinematics in spatially flat FLRW space-times}
\author{Ion I. Cot\u aescu\\ {\small \it  West 
                 University of Timi\c soara,}\\
   {\small \it V. P\^ arvan Ave. 4, RO-1900 Timi\c soara, Romania}}

\maketitle

\begin{abstract}
The kinematics on spatially flat FLRW space-times is presented for the first time  in  co-moving local charts with physical coordinates, i. e. the cosmic time and Painlev\' e-type Cartesian space coordinates.  It is shown that there exists a conserved momentum which determines the form of the covariant four-momentum on geodesics in terms of physical coordinates. Moreover, with the help of the conserved momentum one identifies the peculiar momentum separating the peculiar and  recessional motions without ambiguities. It is shown that the energy and peculiar momentum satisfy the mass-shell condition of special relativity while the recessional momentum does not produce energy. In this framework, the measurements of the kinetic quantities along geodesic performed by different observers are analysed pointing out an energy loss of the massive particles similar to that giving the photon redshift.  The examples of the kinematics on the de Sitter expanding universe and a new Milne-type space-time are extensively analysed.

Pacs: 04.62.+v
\end{abstract}


\newpage

\section{Introduction}

The geodesic motion in general relativity can be described in various local charts (called here frames) as each observer may choose his proper frame with preferred coordinates. From the point of view of general relativity all these frames are equivalent as their coordinates are related through diffeomorphisms under which the mathematical objects transform covariantly \cite{SW,BD}. However, the general diffeomorphisms  are not able to produce conserved quantities such that  we must focus mainly on the isometries which may give rise to conserved quantities via Noether theorem \cite{CML,ES,CEPL}. Thus we must restrict ourselves to a class of observer's frames related through isometries where we have to apply the methods of special relativity in studying the relative motion but using the isometry group instead of the Poincar\' e one as in our recent de Sitter relativity \cite{CdSR1,CdSR2}.

Under such circumstances, it is crucial to understand  which are  the significant physical quantities  and  how these may be related to the conserved quantities generated by isometries.  Another problem is how the measurement depends on the choice of observer's frame taking into account that the isometries transform the conserved quantities among themselves such that different observers measure different values of these quantities. In this paper we would like to analyse these problems in the simple case of the spatially flat Friedmann-Lema\^ itre-Robertson-Walker (FLRW) space-times where we focus on the kinetic quantities on time-like or null geodesics in frames with physical coordinates. 

Apparently these  are trivial problems, that might be solved from long time, but in fact these are still actual since the physical coordinates, giving directly the physical distances,  are those of  Painlev\' e type which were used in static problems but never in the case of the co-moving frames of the spatially flat FLRW space-times. Here we introduce these coordinates obtaining the physical co-moving frames with a time-dependent metric but with spatially flat sections whose Cartesian coordinates give the physical distances as in Minkowski spece-time. We assume that the measured quantities are the components $p^{\mu}$  of the covariant energy-momentum four-vector in the physical frames,  formed by {\em energy}, $p^0$ and {\em covariant} momentum, ${\bf p}$, bearing in mind that, in general, these are functions of cosmic time.  

In other respects,  the spatially flat FLRW space-times have the Euclidean isometry group $E(3)$ formed by space rotations and space translations giving a conserved  angular momentum and a conserved momentum, ${\bf P}$. The angular momentum is related to the symmetry under rotations which is global as we use Cartesian coordinates. The conserved momentum  which does not coincide with the  covariant one is  important  since this generates three prime integrals helping us to derive the energy and covariant momentum we need for integrating the geodesic equation which will be determined by the initial condition and its conserved momentum. Moreover, the conserved momentum helps us to separate the peculiar momentum, proportional with ${\bf P}$, from the recessional one finding that the energy and peculiar momentum in the physical co-moving frame satisfy the mass-shell condition of special relativity. All these results concerning the kinematics in co-moving physical frames are presented in the first part on the next section.

It remain to find how different observers measures the covariant energy and covariant momentum as long as in the FLRW geometries under considerations the translations are isometries transforming the covariant four-vectors and the conserved quantities. In the last part of the next section we investigate how these quantities are measured by two observers staying in two different points of the same geodesic pointing out that, in contrast with the Minkowski space-time, the observer position determines the form and the meaning of the measured quantities. Thus we deduce that the massive or massless particles lose some energy during propagation  which in the massless case is related to the redshift.  

The third section is devoted to a well-known example, namely the de Sitter expanding universe whose geodesics we studied in different frames including  the physical one but without paying attention to the energy and covariant momentum \cite{CdSG}.  The de Sitter manifold has ten independent conserved Killing vectors which generate conserved quantities among them we extract the conserved momentum relating thus the conserved quantities to the measured ones for understanding the role of the conserved quantities in the de Sitter kinematics. The conclusion is that the conserved energy coincides with the measured energy in some points of geodesics while the other conserved quantities, including the conserved momentum, work together for closing the mass-shell condition. The mentioned problem of two observers is also discussed for time-like and null geodesics pointing out the energy loss and redshift.

A new example whose kinematics was never studied is presented in Sec. 4. This is a spatially flat FRLW space-time with  a Milne-type scale factor which, in contrast to the genuine Milne universe, has gravitational sources determining its expansion. We inspect briefly the kinematics on this manifold observing that this behave somewhat complementary with respect to the de Sitter one. Finally we present some concluding remarks.  

In what follows we use the Planck natural units and denote the conserved quantities with capital letters.

\section{Spatially flat FLRW space-times}

The FLRW space-times are the most plausible models of our universe in various periods evolution. The actual universe is observed as being  spatially flat with a reasonable accuracy.  For this reason we focus here on these manifolds for which we consider many types of co-moving frames with Cartesian or spherical coordinates looking for measurable quantities expressed in terms of physical coordinates of Painlev\' e type.   

\subsection{Physical frames}

The Painlev\' e - Gullstrand coordinates \cite{Pan,Gul} were  proposed for studying the Schwarzschild black holes.  Similar coordinates can be introduced in any isotropic manifold $(M,g)$ having  frames $\{x\}=\{t,{\bf x}\}$ with flat space sections. In these frames the coordinates, $x^{\mu}$ ($\alpha,\mu,\nu,...=0,1,2,3$) may be formed by the cosmic time $t$ and either Cartesian space coordinates ${\bf x}=(x^1,x^2,x^3)$ or associated spherical ones $(r,\theta,\phi)$ with Euclidean metric $ds^2_E=d{\bf x}\cdot d{\bf x}=dr^2+r^2 d\Omega^2 $ where $d\Omega^2=d\theta^2+\sin^2\theta\, d\phi^2$. For example, the line element 
\begin{eqnarray}
ds^2=f(r)\, dt_s^2-\frac{dr^2}{f(r)}-r^2 d\Omega^2\,,\label{s1s}
\end{eqnarray}
of any static frame, $\{t_s,r,\theta,\phi\}$, with static time $t_s$, can be put in   Painlev\' e-Gullstrand form, 
\begin{eqnarray}
ds^2=f(r)dt^2+2\sqrt{1-f(r)}\,dtdr-dr^2-r^2 d\Omega^2\,,\label{ss}
\end{eqnarray}
substituting in Eq. (\ref{s1s}) 
\begin{equation}\label{subs}
t_s=t+\int dr \frac{\sqrt{1-f(r)}}{f(r)}\,,
\end{equation}
where $t$ is the {cosmic} time.  

Similar coordinates,  we call here simply Painlev\'e or {\em physical} coordinates, can be introduced in any spatially flat FLRW space-time starting with the conformal Euclidean co-moving frame $\{t_c,{\bf x}_c\}$ with the line element
\begin{equation}\label{conf}
ds^2=a(t_c)^2\left(dt_c^2-d{\bf x}_c\cdot d{\bf x}_c\right)\,.
\end{equation}
Here we may substitute the physical  coordinates $\{t,{\bf x}\}$ defined as
\begin{equation}\label{subs1}
t=\int a(t_c)dt_c\,,  \quad {\bf x}=a(t_c){\bf x}_c\,.
\end{equation}
obtaining the new line element
\begin{equation}\label{Pan}
ds^2=\left(1-\frac{\dot a(t)^2}{a(t)^2}\, {{\bf x}}^2\right)dt^2+2\frac{\dot a(t)}{a(t)}\, {\bf x}\cdot d{\bf x}\,dt -d{\bf x}\cdot d{\bf x}\,, 
\end{equation}
where $a(t)=a[t_c(t)]$ is the usual FRLW scale factor while
\begin{equation}
\frac{\dot a(t)}{a(t)}=\frac{1}{a(t)}\frac{d a(t)}{dt}=\frac{1}{a(t_c)^2}\frac{d a(t_c)}{dt_c}\,,
\end{equation}
is the Hubble function for which we do not use a special notation.  The inverse transformation $\{t,{\bf x}\}\,\to\,\{t_c,{\bf x}_c\}$ is obvious
\begin{equation}\label{subs2}
t_c=\int \frac{dt}{a(t)}\,,  \quad {\bf x}_c=\frac{{\bf x}}{a(t)}\,.
\end{equation}
We suppose that the function $a(t)$ is smooth such that the transformations (\ref{subs1}) and (\ref{subs2}) are diffeomorpkisms.

The metric (\ref{Pan}) is time-dependent laying out an evolving horizon at $|{\bf x}_h|=\frac{a(t)}{\dot a(t)}$ which makes it less popular despite of the fact that these coordinates are just the physical ones, namely the cosmic time $t$ and the Cartesian physical space coordinates ${\bf x}$. Another advantage of this metric is that this is approaching to the Minkowski one  in a neighbourhood of ${\bf x}=0$. 

In many applications one prefers the FLRW coordinates  $\{t,{\bf x}_c\}$ with the well-known line element
\begin{equation}
ds^2=dt^2-a(t)^2 d{\bf x}_c\cdot d{\bf x}_c\,,
\end{equation}
pointing out occasionally the physical distances by multiplying the coordinates with the scale factor $a(t)$. For avoiding this artifice we forget here the FLRW coordinates using directly the physical Painlev\' e coordinates and resorting to the conformal ones as an auxiliary tool when these offer technical advantages.

\subsection{Kinematics}

Our principal objective is to derive the equation of the time-like and null geodesics as well as the associated kinetic quantities  in the physical co-moving frame  $\{t,{\bf x}\}_{O}$ which is the proper frame of an observer staying at rest in the origin $O$. We look for  the components $p^{\mu}=\frac{dx^{\mu}}{d\lambda}$ of the covariant four-momentum  $(p^0, {\bf p} )$ formed by the measured energy $p^0$ and covariant momentum ${\bf p}$. Here  $\lambda$ is an afine parameter related as $ ds=m\,d\lambda$ to the mass  $m$  of a particle moving freely along a geodesic.

We start with an intermediate step, focusing first on the components $p^{\mu}_c=\frac{dx^{\mu}_c}{d\lambda}$ of the  covariant momentum  in the conformal co-moving frame $\{t_c,{\bf x}_c\}_{O}$ of our observer where we have the simple prime integral, 
\begin{equation}\label{int1}
a(t_c)^2\left[p_c^0(t_c)^2-{\bf p}_c(t_c)^2\right]=m^2\,,
\end{equation}
resulted from the line element (\ref{conf}). In other respects, we may exploit the fact that the spatially flat FLRW space-times have $E(3)$ isometries formed by global rotations, $x^i \to R^I_j x^j$ ($i,j,k,...=1,2,3$), and space translations
\begin{equation}\label{trap}
\begin{array}{lll}
t_c&=&t_c'\,,\\
{x}_c^i&=&{x}_c^{\prime\,i}+\xi^i\,,
\end{array} ~~\to~~
\begin{array}{lll}
t&=&t'\,,\\
{x}^i&=&{x}^{\prime\,i}+{\xi}^i \,a(t)\,,
\end{array}
\end{equation}
whose associated Killing vectors $k_{(i)}$ have the components $k^0_{(i)}=0$ and $k^j_{(i)}=\delta_{ij}$ in the frame $\{t_c,{\bf x}_c\}_{O}$ giving rise  to  the conserved quantities 
\begin{equation}\label{Pci}
P^i=- k_{(i)\,j}\frac{dx_c^j}{d\lambda}=a(t_c)^2\frac{dx_c^j}{d\lambda}\,,
\end{equation} 
representing  the components of the {\em conserved} momentum which is different from the covariant momentum ${\bf p}(t)$. Then by using the prime integrals (\ref{int1}) and (\ref{Pci}) we derive the energy and covariant momentum  in this frame as
\begin{eqnarray}
p_c^0(t_c)&=&\frac{dt_c}{d\lambda}=\frac{1}{a(t_c)} \sqrt{m^2+\frac{{P}^{2}}{a(t_c)^2}}\,,\label{uu0}\\
p_c^i(t_c)&=&\frac{d{x_c^i}}{d\lambda}= \frac{P^i}{a(t_c)^2}\,,\label{uui}
\end{eqnarray}
where  we denote $P=|{\bf P}\,|$. The geodesic results simply  as
\begin{equation}
\frac{dx_c^i}{dt_c}=\frac{p_c^i(t_c)}{p^0_c(t_c)} ~~\to~~x^i_c(t_c)=x_{c0}^i +\frac{P^i}{m}\int_{t_{c0}}^{t_c}\frac{dt_c}{\sqrt{a(t_c)^2+\frac{P^2}{m^2}}}\,,
\end{equation}
concluding that any time-like geodesic is determined completely by its conserved momentum ${\bf P}={\bf n}_P P$  and the initial condition ${\bf x}_c(t_{c0})={\bf x}_{c0}$. This equation must be integrated in each particular case but for the massless particles (with $m=0$) we have the universal solution
\begin{equation}\label{null1}
{\bf x}_c(t_c)={\bf x}_{c0}+{\bf n}_P(t_c-t_{c0})\,,
\end{equation}
giving the null geodesics on any FLRW space-time.

The corresponding physical quantities measured by the observer $O$ in his physical proper frame, $\{t,{\bf x}\}_{O}$, may be obtained by substituting the physical coordinates according to Eq. (\ref{subs2}). Thus we find the covariant components,
\begin{eqnarray}
p^0(t)&=&\frac{dt}{d\lambda}= \sqrt{m^2+\frac{{P}^{2}}{a(t)^2}}\,,\label{uu01}\\
p^i(t)&=&\frac{d{x^i}}{d\lambda}= \frac{P^i}{a(t)}+x^i(t)\,\frac{\dot a(t)}{a(t)}\sqrt{m^2+\frac{{P}^{2}}{a(t)^2}}\,,\label{uui1}
\end{eqnarray}
which represent the measured energy and covariant momentum in the point $[t, {\vec x}(t)]$ of the time-like geodesic 
\begin{equation}\label{geod}
{\bf x}(t)={\bf x}_0\,\frac{a(t)}{a(t_0)}+ \frac{\bf P}{m}\,a(t)\int_{t_0}^{t}\frac{dt}{a(t)\sqrt{a(t)^2+\frac{P^2}{m^2}}}
\end{equation}
which is passing through the space point ${\bf x}(t_0)={\bf x}_0$ at the initial time $t_0$. In the physical frame $\{t,{\bf x}\}_{O}$ the equation of the null geodesics,
\begin{equation}\label{null2}
{\bf x}(t)={\bf x}_0\,\frac{a(t)}{a(t_0)}+ {\bf n}_P\, a(t)\left[t_c(t)-t_c(t_0)\right]\,,
\end{equation}
results from  Eq. (\ref{null1}).  

A special problem is that of tachyons whose kinetic quantities on space-like geodesics can be obtained by substituting $m^2\,\to\,-m^2$ in the above equations. Then the  energy
\begin{equation}
p^0_{\rm tach}(t)=\sqrt{-m^2+\frac{{P}^{2}}{a(t)^2}}\,,
\end{equation}
is real valued only when $a(t)<\frac{P}{m}$. This means that in expanding universes a tachyon with conserved momentum ${\bf P}$ disappears when  $a(t)$ reaches the value $\frac{P}{m}$ this surviving only in collapsing universes for smaller values of the scale factor. As here we focus only on expanding geometries we ignore the space-like geodesics remaining to study the time-like and null ones.

The momentum defined by Eq. (\ref{uui1}) can be split  as  ${\bf p}(t)=\hat{\bf p}(t)+ \bar{\bf p}(t)$ where
\begin{equation}\label{ppx}
\hat {\bf p}=\frac{\bf P}{a(t)}\,,\quad \bar{\bf p}={\bf x}(t) \,\frac{\dot a(t)}{a(t)}\, p^0(t)\,,
\end{equation}
are the {\em peculiar} and respectively {\em recessional} momenta. The prime integral derived from the line element (\ref{Pan}) gives the familiar identity
\begin{equation}\label{Pan1}
p^0(t)^2- \hat{\bf p}(t)^2=m^2\,, 
\end{equation}
which is just the mass-shell condition of special relativity satisfied by the energy and peculiar momentum along the geodesic. Thus we see that in the physical co-moving frame the peculiar momentum can be separated naturally being proportional with the conserved momentum. Moreover, this produces the entire energy of the geodesic as in special relativity. Thus for  ${\bf P}=0$ and $p^0(t)=m$, the particle remaining at rest in the point  ${\bf x}(t)=\frac{a(t)}{a(t_0)}{\bf x}_0$ but  moving with the recessional momentum $\bar{\bf p}$ with respect the observer $O$.  We must stress that these properties hold only in the co-moving frames with physical coordinates since in other frames this separation is not possible while the momenta satisfy dispersion relations depending explicitly on time as in Eq. (\ref{int1}) or identity $p^0(t)^2-a(t)^2 {\bf p}_c(t)^2=m^2$   that holds in FLRW coordinates $\{t,{\bf x}_c\}$.

Hereby other interesting kinetic quantities can be derived as, for example, the velocity 
\begin{equation}\label{vel}
{\bf v}(t)=\frac{{\bf p}(t)}{p^0(t)}={\bf x}(t)\,\frac{\dot a(t)}{a(t)}+ \frac{\hat{\bf p}(t)}{\sqrt{m^2+\hat {\bf p}(t)^2)}}\,,
\end{equation}
whose first term is the recessional velocity due to the space evolution, complying with the velocity law which is confused sometimes with the Hubble one \cite{H0,H}. The second term is the peculiar velocity which depends on the peculiar momentum as in special relativity. 
 
The covariance under rotations, which behave here as a global symmetry, gives rise to the conserved angular momentum, that depends only on the peculiar momentum
\begin{equation}
{\bf L}={\bf x}(t)\land {\bf p}(t)={\bf x}(t)\land\hat {\bf p}(t)=\frac{{\bf x}(t)\land {\bf P}}{a(t)}=\frac{{\bf x}_0\land {\bf P}}{a(t_0)}\,,
\end{equation}
and can be related to the initial condition. This vanishes when the observer $O$ stays at rest in a space point of the measured geodesic.

\subsection{Two observers problem}

The physical quantities ${p^0}(t)$, $\hat {\bf  p}(t)$  and $\hat{\bf  p}(t)$ are functions of time but the last one depends explicitly on coordinate such that  the experimental results will depend on the relative position between the detector and the measured particle.  However, this is not an impediment as the peculiar and recession contributions can be separated at any time without ambiguities. Nevertheless, for avoiding confusions  we take care on this dependence  looking for suitable positions of  observer's frames in order to obtain intuitive results.  

The example we would like to discuss here is of two observers  measuring the motion of a massive particle on a time-like geodesic which is passing through the origins $O$ and $O'$ of their proper co-moving frames $\{t,{\bf x}\}_{O}$ and $\{t,{\bf x}'\}_{O'}$.  We assume that at the initial time $t_0$ the origin  $O'$ is translated with respect to $O$ as ${\bf x}(t_0)={\bf x}'(t_0)+{\bf d}(t_0)$ where ${\bf d}(t_0)={\bf d}\,a(t_0)$ depends on   the translation parameters of Eqs. (\ref{trap}) denoted now by $\xi^i=d^i$. Then it is convenient to introduce the unit vector ${\bf n}$ of the direction $OO'$ such that ${\bf d}={\bf n}\, d$.

Our experiment  starts in this lay out  at the time $t_0$ when the observer $O'$ lunches a particle of mass $m$,  momentum ${\bf p}=-{\bf n}\, p$ and energy $p^0=\sqrt{m^2+p^2}$ on the geodesic $O'\to O$. The problem is to find which are the energy and momentum of this particle  measured in the origin $O$  at the final time $t_f$ when the particle reach this point.  For solving this problem we look first for the conserved momentum that can be derived in $O'$ as 
\begin{equation}\label{Pn}
{\bf P}={\bf n}_P P={\bf p}\,a(t_0)~ ~\to~~P=p\, a(t_0)\,,~~~{\bf n}_P=-{\bf n}\,.
\end{equation}
Then by using Eqs. (\ref{uu01}) and (\ref{uui1}) we obtain the momentum and energy measured in $O$, 
\begin{eqnarray}
p^0(t_f)&=& \sqrt{m^2+p^2 \frac{a(t_0)^2}{a(t_f)^2}}\,,\label{pef}\\
{\bf p}(t_f)&=&\hat{\bf p}(t_f)= -{\bf n}\, p\,\frac{a(t_0)}{a(t_f)}\,,\label{mef} 
\end{eqnarray}
where $t_f$ is the solution of the equation ${\bf n}\cdot{\bf x}(t_f)=0$ with ${\bf x}(t)$ given by Eq. (\ref{geod}). This equation can be written simply as 
\begin{equation}\label{geodOO}
\frac{P}{m}\int_{t_0}^{tf}\frac{dt}{a(t)\sqrt{a(t)^2+\frac{P^2}{m^2}}}=d
\end{equation}
where  $d$ is the time-independent translation parameter defined above. 
Solving Eq. (\ref{geodOO}) we obtain a function $t_f (P,t_0)$ which must be singular in $P=0$ for preventing the left handed term of this equation on vanishing in this limit.
Once we have the value of $t_f$ we can calculate the propagation time $t_f-t_0$,  the distance  $d(t_f)$ between $O$ and $O'$ at the time $t_f$ and the final peculiar velocity $\hat v(t_f)$ of the particle arriving in $O$. According to Eqs. (\ref{trap}) and (\ref{pef}) we find
\begin{eqnarray}
d(t_f)&=&d\, a(t_f)=d(t_0)\,\frac{a(t_f)}{a(t_0)}\,,\label{dtf}\\
\hat  v(t_f)&=&\left( 1+\frac{m^2}{p^2}\frac{a(t_f)^2}{a(t_0)^2} \right)^{-\frac{1}{2}}\,,\label{vtf}
\end{eqnarray} 
completing thus the collection of kinetic quantities related to this problem. 

Eq. (\ref{pef})  shows that in expanding universes a part of energy is lost during the propagation. This can be measured by the relative energy loss defined as
\begin{equation}
e=1-\frac{p^0(t_f)}{p^0(t_0)}\,.
\end{equation}
This phenomenon is similar to the redshift of the photons with $m=0$ for which we recover the Lema\^ itre equation \cite{L1,L2} of Hubble's law \cite{Hubb} as 
\begin{equation}\label{red}
\frac{1}{1-e}= 1+z=\frac{p^0(t_0)}{p^0(t_f)}=\frac{a(t_f)}{a(t_0)}\,,
\end{equation}
where $z$ is the usual redshift defined as the relative dilation of the wave length. As was expected for $m=0$ the final velocity  $\hat v(t_f)=1$ is the speed of light.   

All the results presented here  can be exploited effectively only in concrete geometries where the geodesic equation can be integrated. In what follows we discuss two such examples starting with one of the most studied geometries.  
 
\section{de Sitter expanding universe}

The first example is the expanding portion of the de Sitter space-time defined as the hyperboloid of radius $1/\omega_H$  in the five-dimensional flat spacetime $(M^5,\eta^5)$ of coordinates $z^A$  (labelled by the indices $A,\,B,...= 0,1,2,3,4$) having the metric $\eta^5={\rm diag}(1,-1,-1,-1,-1)$. The coordinates $\{x\}$   can be introduced giving the set of functions $z^A(x)$ which solve the hyperboloid equation,
\begin{equation}\label{hip}
\eta^5_{AB}z^A(x) z^B(x)=-\frac{1}{\omega_H^2}\,.
\end{equation}
where  $\omega_H$ is the Hubble de Sitter constant  in our notations.  

There are co-moving frames with conformal coordinates $\{t_c,{\bf x}_c\}$ or with physical ones $\{t,{\bf x}\}$ having the scale factors
\begin{equation}
a(t)=e^{\omega_H t}~\to~ t_c=-\frac{1}{\omega_H}e^{-\omega_H t}~\to~ a(t_c)=-\frac{1}{\omega t_c}\,,
\end{equation} 
defined for $t\in {\Bbb R}$ and $t_c<0$ corresponding to the expanding portion. In this case the Hubble function becomes the constant $\omega_H$. In addition, this manifold allows even a static frame $\{t_s,{\bf x}\}$ with the line element (\ref{s1s}) where $f(r)=1-\omega_H^2 r^2$ and $t_s=t-\ln f(r)$.  

\subsection{Conserved quantities}

The de Sitter manifold has a rich  isometry group which is just the gauge group $SO(1,4)$ of the embedding manifold $(M^5,\eta^5)$  that leave  invariant its metric and implicitly Eq. (\ref{hip}). Therefore, given a system of coordinates defined by the functions $z=z(x)$, each transformation ${\frak g}\in SO(1,4)$ defines the isometry $x\to x'=\phi_{\frak g}(x)$ derived from the system of equations
\begin{equation}\label{zz}
z[\phi_{\frak g}(x)]={\frak g}z(x)
\end{equation}
that holds for any type of coordinates which means that  these isometries are defined globally. The sets of local charts related through these isometries play the role of the   inertial frames similar to those of special relativity. 

Given an isometry $x\to x'=\phi_{{\frak g}(\xi)}(x)$ depending on the group parameter $\xi$ there exists an associated Killing vector, ${k}=\partial_{\xi}\phi_{\xi}|_{\xi=0}$ (which satisfy the Killing equation ${k}_{\mu;\nu}+{k}_{\nu;\mu}=0$). Thus in a canonical parametrization of the $SO(1,4)$ group, with real skew-symmetric parameters $\xi^{AB}=-\xi^{BA}$, any infinitesimal isometry can be written as
$\phi^{\mu}_{{\frak g}(\xi)}(x)=x^{\mu}+ \xi^{AB}k^{\mu}_{(AB)}(x)+...$.   
Starting with the general definition of the Killing vectors in the pseudo-Euclidean spacetime $(M^5,\eta^5)$, we may consider the following identity
\begin{equation}
K^{(AB)}_Cdz^C=z^Adz^B-z^Bdz^A=k^{(AB)}_{\mu}dx^{\mu}\,,
\end{equation}
giving the covariant components  of the Killing vectors in an arbitrary frame $\{x\}$ of  the de Sitter manifold as
\begin{equation}\label{KIL}
k_{(AB)\,\mu}=\eta^5_{AC}\eta^5_{BD}k^{(CD)}_{\mu}= z_A\partial_{\mu}z_B-z_B\partial_{\mu}z_A\,, 
\end{equation}
where $z_A=\eta_{AB}z^B$. 

The classical conserved quantities  along the time-like geodesics  have the general form  ${\cal K}_{(AB)}(x,{\bf P})=\omega_H  k_{(AB)\,\mu} p^{\mu}$ where $p^{\mu}$  are the components of the covariant four-vector defined above. The conserved quantities with physical meaning \cite{CGRG} are, the energy $E=\omega_H  k_{(04)\,\mu}p^{\mu}$, the angular momentum components,  $L_i= \frac{1}{2}\,\varepsilon_{ijk}k_{(jk)\,\mu}p^{\mu}$, and the components $K_i=  k_{(0i)\,\mu} p^{\mu}$ and $R_i=  k_{(i4)\,\mu} p^{\mu}$
of two vectors  related to  the conserved momentum ${\bf P}$ and its associated dual momentum ${\bf Q}$ as,  
\begin{equation}\label{PQKR}
{\bf P}=-\omega_H ({\bf R}+{\bf K})\,, \quad {\bf Q}=\omega_H ({\bf K}-{\bf R})\,. 
\end{equation}
satisfying the identity
\begin{equation}\label{disp}
E^2-\omega_H^2 {{\bf L}}^2-{\bf P}\cdot {\bf Q}=m^2\,,
\end{equation}
corresponding to the first Casimir invariant of the $so(1,4)$ algebra \cite{CGRG}. In the flat limit, $\omega_H\to 0$  and $-\omega_H t_c\to 1$, we have ${\bf Q} \to {\bf P}$ such that this identity  becomes just  the usual null mass-shell condition $E^2-{\bf P}^2=m^2$ of  special relativity.

Note that the conserved quantities transform among themselves under de Sitter isometries including the simple translations which in this case transform the energy and dual momentum as we have shown recently \cite{CdSR1}. 

\subsection{Time-like geodesics}

The coordinates of the physical co-moving  frame $\{t,{\bf x}\}_{O}$  are introduced by the functions 
\begin{eqnarray}
z^0(x)&=&\frac{1}{2\omega_H}\left[e^{\omega_H t}-e^{-\omega_H t}(1 - \omega_H^2{\bf x}^2)\right]\,,
\nonumber\\
z^i(x)&=&x^i \,, \label{Zx}\\
z^4(x)&=&\frac{1}{2\omega_H}\left[e^{\omega_H t}+e^{-\omega_H t}(1 - \omega_H^2{\bf x}^2)\right]\,.
\nonumber
\end{eqnarray}
giving the line element 
\begin{equation}\label{mdSP}
ds^2=(1-\omega_H^2 {{\bf x}}^2)dt^2+2\omega_H {\bf x}\cdot d{\bf x}\,dt -d{\bf x}\cdot d{\bf x}\,, 
\end{equation}
having the horizon at $|{\bf x}_h|=\omega_H^{-1}$ such that the condition $\omega_H|{\bf x}|<1$ is mandatory.
In this frame the equation of a time-like geodesic  can be obtained by solving the integral of Eq. (\ref{geod}). We obtain thus the geodesic equation   \cite{CdSG}, 
\begin{eqnarray}
{\bf x}(t) &=&{\bf x}_0e^{\omega_H( t-t_0)}+{\bf n}_P\frac{e^{\omega_H t}}{\omega_H P}\left(\sqrt{m^2+P^2e^{-2\omega_Ht_0}}\,\right.\nonumber\\
&&\hspace*{38mm}-\left.\sqrt{m^2 +P^2e^{-2\omega_H t}}\right)\,,\label{geodS}
\end{eqnarray}
determined by the conserved momentum ${\bf P}={\bf n}_P P $  and the initial condition ${\bf x}(t_0)={\bf x}_0$ fixed at the time $t_0$. 

The conserved quantities  in an arbitrary point $(t,{\bf x}(t))$  of this geodesic can be expressed as  \cite{CdSR1,CdSG},
\begin{eqnarray}
E&=&\omega_H\, {\bf x}(t)\cdot {\bf P}\, e^{-\omega_H t}+\sqrt{ m^2+{P}^{2}e^{-2\omega_Ht}}\,,\label{Ene}\\
{\bf L}&=& {\bf x}(t)\land {\bf P}e^{-\omega_H t}\,,\label{La}\\
{\bf Q}&=&2\omega_H\, {\bf x}(t)E e^{-\omega_H t}+{\bf P}  e^{-2\omega_H t}[1-\omega_H^2{\bf x}(t)^2]\,.\label{Qa}
\end{eqnarray}
satisfying the  identity (\ref{disp}). Moreover, the Eqs. (\ref{uu01}) and (\ref{uui1}) give the energy and covariant momentum components,
\begin{eqnarray}
p^0(t)&=&\frac{dt}{d\lambda}=\sqrt{m^2 +P^2 e^{-2\omega_Ht}}\,,\\
p^i(t) &=&\frac{dx^i}{d\lambda}=e^{-\omega_Ht}P^i+\omega_H x^i(t) \sqrt{m^2 +P^2 e^{-2\omega_Ht}}\,,
\end{eqnarray} 
that can be measured by the observer $O$ in his proper frame  $\{t,{\bf x}\}_{O}$. 
The conserved quantities are related to the measured ones as 
\begin{eqnarray}
E&=&\omega_H\, {\bf x}(t)\cdot \hat{\bf p}(t)+p^0(t)\,,\label{Ep}\\
{\bf L}&=& {\bf x}(t)\land \hat{\bf p}(t)\,,\label{Lp}\\
{\bf P}&=&\hat{\bf p}(t)\,e^{\omega_H t}\label{Pp}\\
{\bf Q}&=&e^{-\omega_H t}\left\{2\omega_H\, {\bf x}(t)E +\hat{\bf p}(t)  [1-\omega_H^2{\bf x}(t)^2]\right\}\,.\label{Qp}
\end{eqnarray}
Hereby we conclude that the conserved quantities depend only on position and peculiar momentum. Among them only $E$ and ${\bf L}$ can be measured while ${\bf P}$ and ${\bf Q}$ are not accessible directly, their role consisting only in closing the invariant (\ref{disp}) as 
\begin{equation}
E^2-\omega_H^2 {\bf L}^2-{\bf P}\cdot {\bf Q}=p^0(t)^2-\hat{\bf p}(t)^2=m^2\,.
\end{equation}
For example, a measurement in observer's origin $O$ gives $E=p^0$, ${\bf L}=0$,  ${\bf P}=\hat{\bf p}\,e^{\omega_H t}$ and ${\bf Q}=\hat{\bf p}\,e^{-\omega_H t}$ such that ${\bf P}\cdot{\bf Q}=\hat{\bf p}^2$. Note that there is a natural choice of the initial moment, $t_0=0$, for which we have $\hat{\bf p}(t_0)={\bf P}={\bf Q}$  and the calculations become simpler. 

Now we can revisit the problem of Sec. 2.3 looking for the value of $t_f$  which solves the equation (\ref{geodOO}). Taking into account that now ${\bf P}={\bf p}\,e^{\omega_H t_0}=-{\bf n}\, p\,e^{\omega_H t_0} $ and ${\bf x}_0={\bf d}(t_0) ={\bf n} d(t_0)$ we obtain the identity
\begin{equation}\label{bau}
\frac{a(t_0)^2}{a(t_f)^2}=e^{-2\omega_H(t_f-t_0)}=\frac{1}{p^2}\left({p^0}-\omega_H d(t_0) p\right)^2-\frac{m^2}{p^2}\,,
\end{equation}
which may be substituted in Eq. (\ref{pef}) leading to the final result
\begin{eqnarray}
p^0(t_f)&=&p^0-\omega_H {d}(t_0){p}=p^0+\omega_H {\bf d}(t_0)\cdot {\bf p}\,,\\
{\bf p}(t_f)&=&\hat{\bf p}(t_f)=-{\bf n}\sqrt{p^0(t_f)^2-m^2}\,,
\end{eqnarray}
expressed exclusively in terms of physical quantities. Hereby we deduce the relative energy loss 
\begin{equation}
e=\omega_H d(t_0)\frac{p}{p^0}=\omega_H d(t_0) v\,,
\end{equation}
proportional with the initial velocity $v$ of the particle lunched by $O'$. In the case of the massless photons $v=1$ recovering the energy loss producing the redshift.  It remains to derive  the final distance and velocity which take the form
\begin{eqnarray}
d(t_f)&=&d(t_0)\,\frac{p}{|{\bf p}(t_f)|}\,,\label{dtf1}\\
\hat  v(t_f)&=&\left( 1+\frac{m^2}{|{\bf p}(t_f)|^2} \right)^{-\frac{1}{2}}\,,\label{vtf1}
\end{eqnarray} 
as it results from Eqs. (\ref{dtf}), (\ref{vtf})  and (\ref{bau}).

For understanding the role of the conserved quantities in this experiment we must specify  that  the observers $O$ and $O'$ record different conserved quantities since the translation is an isometry which changes the components of the conserved quantities apart from the conserved momentum which is not affected by these isometries \cite{CdSR1}. Moreover, as the origins of these frames are on the geodesics, both the angular momenta measured in $O$ and $O'$ vanishes. We denote by $E, {\bf Q}$ the remaining conserved quantities measured in $O$ and by  $E', {\bf Q}'$ those recorded in $O'$ bearing in mind that 
\begin{equation}
{\bf P}'={\bf P}={\bf p}\,e^{\omega_H t_0}\,.
\end{equation}
The values observed in $O'$ can be deduced from Eqs.  (\ref{Ene}) and (\ref{Qa}) for ${\bf x}'=0$ obtaining the previous mentioned result, $E'=p^0$ and ${\bf Q}'={\bf p}\,e^{-\omega_H t_0}$. The observer $O$ prefers to look for the conserved quantities at the time $t_0$  since his knows that these do not change along the geodesic. Thus he records
\begin{eqnarray}
E&=&p^0(t_f)\,,\\
{\bf Q}&=&e^{-\omega_H t_0}\left[2\omega_H {\bf d}(t_0) E+{\bf p}\left( 1-\omega_H^2 {\bf d}(t_0)^2\right)\right]\,,
\end{eqnarray}
as it results from Eqs.  (\ref{Ene}) and (\ref{Qa}) for ${\bf x}_0={\bf d}(t_0)$,   verifying  that  ${\bf P}\cdot{\bf Q}={\bf p}(t_f)^2$ for closing again the identity (\ref{disp}). Note that the relation among the conserved quantities $E, {\bf P},...$ and 
$E', {\bf P}',...$ can be derived directly according to the transformation rule under isometries we have discussed recently \cite{Dop1,Dop2}.

\subsection{Null geodesics}

The de Sitter null geodesics of the photons with $m=0$  that read 
\begin{equation}
{\bf x}(t)={\bf x}_0\, e^{\omega_H(t-t_0)}+{\bf n}_{P}\frac{e^{\omega_H(t-t_0)}-1 }{\omega_H}\,,
\end{equation}
are interesting being involved in the theory of the redshift.  The energy and covariant momentum denoted now by $k^0(t)$ and respectively ${\bf k}(t)$ are
\begin{eqnarray}
k^0(t)&=&P\,e^{-\omega_H t}=|\hat{\bf k}(t)|\,,\\
{\bf k}(t)&=&e^{-\omega_H t}{ P}\left({\bf n}_P+\omega_H{\bf x}(t)\right)=\hat{\bf k}(t)+\bar{\bf k}(t)
\end{eqnarray} 
such that we can separate the peculiar momentum,  $\hat{\bf k}(t)=e^{-\omega_H t}{\bf P}$, and  the recessional one, $\bar{\bf k}(t)=\omega_H {\bf x}(t) P e^{-\omega_H t}=\omega_H {\bf x}(t) k^0(t)$. 

Considering again the problem of Sec. 2.3 we assume that now the observer $O'$  emits a photon of energy $k$ and momentum ${\bf k}=-{\bf n}k$. Under such circumstances Eq. (\ref{bau}) gives
\begin{equation}
\frac{a(t_0)}{a(t_f)}=e^{-\omega_H(t_f-t_0)}=k\left[1-\omega_H d(t_0)\right]\,,
\end{equation} 
allowing us to derive the quantities observed by $O$ in his proper frame, namely the energy and covariant momentum,
\begin{eqnarray}
k^0(t_f)&=&k\,\left[1-\omega_H {d}(t_0)\right]\,,\\
{\bf k}(t_f)&=&\hat{\bf k}(t_f)=-{\bf n}\,k^0(t_f)\,,
\end{eqnarray}
the value of the final time 
\begin{equation}
t_f=t_0-\frac{1}{\omega_H}\ln\left[1-\omega_H d(t_0)\right]\,,
\end{equation}
the final distance between $O$ and $O'$ at the time $t_f$, 
\begin{equation}
d(t_f)=\frac{d(t_0)}{1-\omega_H d(t_0)}\,,
\end{equation}
and the redshift $z$ related to the relative energy loss $e$ observed by $O$,  
\begin{equation}\label{redS}
1-e=\frac{1}{1+z}=1-\omega_H d(t_0)\,,
\end{equation}
resulted from Eq. (\ref{red}). We recall that the condition $\omega_H d(t_0)<1$ is mandatory. 

On the null geodesics the conserved quantities have simpler forms as 
\begin{eqnarray}
E&=&k^0(t_f)=k\,\left[1-\omega_H {d}(t_0)\right]\,,\\
{\bf P}&=&{\bf k}\,e^{\omega t_0}\,,\\
{\bf Q}&=&{\bf k}\,e^{-\omega_H t_0}\left[1-\omega_H {d}(t_0)\right]^2\,,
\end{eqnarray}
such that ${\bf P}\cdot {\bf Q}=E^2$ satisfying the identity (\ref{disp}) with $m=0$.
We observe again that for the special choice $t_0=0$ we have ${\bf P}={\bf k}$ and ${\bf d}(t_0)={\bf d}$ which simplifies the calculations and their interpretation. 

\section{Milne-type universe}

Let us finish with an example of manifold whose kinematics was never studied. This is the spatially flat FLRW manifold $M$ with the Milne type scale factor $a(t)=\omega_M t$ defined on the domain  $t\in (0,\infty)$, whose constant (frequency) $\omega_M$ is introduced from dimensional reasons \cite{M1,M2}. Then we may write the line element in the physical co-moving  frame  $\{t,{\bf x}\}$ as
\begin{equation}\label{lineM}
ds^2=\left(1-\frac{1}{t^2}{\bf x}^2\right)dt^2 + 2 {\bf x}\cdot d{\bf x}\,\frac{dt}{t}-d{\bf x}\cdot d{\bf x}\,,
\end{equation}
after substituting in Eq. (\ref{Pan}) the Hubble function $\frac{\dot a(t)}{a(t)}=\frac{1}{t}$ which is independent on $\omega_M$. The conformal time $t_c\in (-\infty,\infty)$ is defined as 
\begin{equation}\label{tt}
t_c=\int \frac{dt}{a(t)}=\frac{1}{\omega_M} \ln(\omega_M t) ~\to~  a(t_c)=e^{\omega_M t_c}\,,
\end{equation}
obtaining the function $a(t_c)$ of the line element (\ref{conf}) of the conformal co-moving frame $\{t_c,{\bf x}_c\}$. 

Here  the constant $\omega_M$  is an useful free parameter representing the expansion speed of $M$. We remind the reader that in the case of the genuine Milne universe (of negative space curvature but globally flat) one must set $\omega_M=1$ for eliminating the gravitational sources \cite{BD}. In contrast, our space-time $M$ is produced by isotropic gravitational sources, i. e. the density $\rho$ and pressure $\underline{p}$, evolving in time as \cite{M1}
\begin{equation}
\rho=\frac{3}{8\pi G}\frac{1}{t^2}\,, \quad \underline{p}=-\frac{1}{8\pi G}\frac{1}{t^2}\,,
\end{equation}
and vanishing for $t\to\infty$. These sources govern the expansion of $M$ that can be better observed in the frame $\{t, {\bf x}\}$ where the line element (\ref{lineM}) lays out an expanding horizon at $|{\bf x}_h|=t$ and tends to the Minkowski space-time when $t\to \infty$ and the gravitational sources vanish.

We deduce first the  equation of the time-like geodesics, solving the integral of Eq. (\ref{geod}), which leads to the final form
\begin{equation}\label{geodM}
{\bf x}(t)=\frac{t}{t_0}\,{\bf x}_0+ {\bf n}_P\, {t}\ln\left( \frac{t}{t_0}\,\frac{P+\sqrt{P^2+\omega_M^2 m^2 t_0^2}}{P+\sqrt{P^2+\omega_M^2 m^2 t^2}} \, \right)\,,
\end{equation}
that for $m=0$ gives the equation 
\begin{equation}
{\bf x}(t)=\frac{t}{t_0}\,{\bf x}_0+ {\bf n}_P\, {t}\ln\left(\frac{t}{t_0}\right)\,,
\end{equation}
of the null geodesics.  The energy, momentum and velocity have to be derived according to Eqs. (\ref{uu01}), (\ref{uui}) and (\ref{vel}). These are complicated formulas but that can be used in applications by using algebraic codes on computer. 

Furthermore, coming back to the problem of two observers formulated in Sec. 2.3.,
we solve Eq. (\ref{geodOO}) for deriving the final time $t_f$ and the ratio
\begin{equation}
\frac{a(t_0)}{a(t_f)}=\frac{t_0}{t_f}=\frac{1}{2}\frac{e^{-\omega_Md}(p_0+p)^2-e^{\omega_M d}\,m^2}{p(p^0+p)}
\end{equation}
We recall that $p$ is the scalar initial momentum of the particle of mass $m$ lunched by $O'$ at the time $t_0$. Then, according to Eqs, (\ref{pef}) and (\ref{mef}) we find the final energy and covariant momentum
\begin{eqnarray}
p^0(t_f)&=&\frac{1}{2}\frac{e^{-\omega_Md}(p_0+p)^2+e^{\omega_M d}\,m^2}{p^0+p}\\
{\bf p}(t_f)&=&-{\bf n}\,\frac{1}{2}\frac{e^{-\omega_Md}(p_0+p)^2-e^{\omega_M d}\,m^2}{p^0+p}\,,
\end{eqnarray} 
and the final distance between $O$ and $O'$ when the particle arrives in $O$,
\begin{equation}
d(t_f)=d(t_0) \frac{2p(p^0+p)}{e^{-\omega_Md}(p_0+p)^2-e^{\omega_M d}\,m^2}\,,
\end{equation}
where $d(t_0)=d\, a(t_0)=\omega_M t_0 d$. As in this geometry the horizon is at $t$ we must impose the restriction $d(t_0)<t_0 \to \omega_M d<1$ such that the final distance remains inside the horizon, $d(t_f)< t_f$.

When $O$ and $O'$ observe a photon then they record $t_f=t_0\, e^{\omega_M d}$, $k^0(t_f)=|{\bf k}(t_f)|=k\, e^{-\omega_M d}$  and the redshift $1+z=e^{\omega_M d}$ which for small values of $\omega_M d$ can be confused with the de Sitter one since the expansion
\begin{equation}
\frac{1}{1+z}=e^{-\omega_M d}=1-\omega_M d +{\cal O}(\omega_M^2 d^2)\,,
\end{equation}
is somewhat similar with Eq. (\ref{redS}). However, for larger distances the discrepancy between the linear behaviour of the de Sitter redshift and the exponential one in the space-time $M$  becomes obvious.

Finally we observe that the Milne-type and ds Sitter universes behave somewhat complementary such that the cosmic time of one of these manifolds behaves as the conformal time of the other one. The self explanatory next table completes this image \cite{M2}.
\begin{center}
\begin{tabular}{ccc}
&$M$&de Sitter\\
&&\\
$t$&$0<t=\frac{1}{\omega_M}e^{\omega_M t_c}<\infty$& $-\infty<t<\infty$\\
$t_c$&$-\infty<t_c<\infty$&$-\infty<t_c=-\frac{1}{\omega_H}e^{-\omega_H t}<0$ \\
$a(t)$&$\omega_M t$&$~~~~e^{\omega_H t}~~~$\\
$a(t_c)$&$e^{\omega_M t_c}$&$-\frac{1}{\omega_H t_c}$\\
transl.&$\omega_M d<1$&$\omega_H d<1$\\
$1+z$&$e^{\omega_M d}$&$[1-\omega_H d(t_0)]^{-1}$
\end{tabular}
\end{center}
The only similarity is the condition satisfied by the translation parameter $d$ for remaining inside the horizon.  

\section{Concluding remarks}

We presented here the complete kinematics in co-moving frames with physical coordinates on spatially flat FLRW space-times, based on the conserved quantities among them the conserved momentum is the central piece of our approach. In these   frames, the geodesics are determined completely by the initial condition and conserved momentum. Moreover, this allows us to separate the peculiar motion from the recessional one such that  the energy and peculiar momentum  satisfy the mass-shell condition of special relativity. In this framework we discussed the problem of two observers pointing out the relative energy loss during propagation which in the massless case gives the well-known redshift. 

The first example is the kinematics of the de Sitter expanding universe related to our previous results concerning the geodesics of this manifold \cite{CdSG}. Here we presented for the first time the measurable quantities on geodesics in physical co-moving frames showing how these are related to the rich set of the conserved quantities of this geometry. We observed that only the conserved energy is related directly to the measured one while the conserved momentum and its dual help each other in closing the mass-shell relation. Moreover, we pointed out that the meaning of the conserved momentum depends on the choice of the initial time showing that we can set this time as the moment in which the conserved momentum coincides with the covariant initial momentum. This observation is important since the momentum operator of de Sitter quantum mechanics is related to the conserved momentum \cite{CGRG}.

The second example we presented here for the first time is the kinematics of a new manifold we considered recently in quantum theory \cite{M1,M2}. This is a spatially flat FLRW space-time with  a Milne type scale factor produced by gravitational sources proportional with $t^{-2}$. The geodesic motion on this manifold was studied in physical co-moving frames deriving the kinetic quantities on geodesics and outlined the results of the experiment of two observers including the redshift. Moreover, we argued that this manifold is interesting since it behaves complementary to the de Sitter one having thus two different  examples of FRLW  kinematics. 

As a final conclusion we may say that the physical coordinates and the conserved momentum offer the suitable framework in which we can distinguish without any ambiguity between the recessional motion due to the background expansion and the peculiar one which behaves just as in special relativity. Thus we may get a new perspective in interpreting the astrophysical measurements in our actual expanding universe.

\end{document}